\documentclass[prd,aps,10pt,twocolumn,floats,floatfix,nofootinbib] {revtex4-1}

\usepackage{graphicx}
\usepackage{dcolumn}
\usepackage{bm}
\usepackage{graphics}
\usepackage{slashed}
\usepackage{amssymb}
\usepackage{natbib}
\usepackage{amsmath}
\usepackage{url}
\usepackage[usenames,dvipsnames,svgnames,table]{xcolor}
\usepackage{color}
\usepackage{xcolor}
\usepackage{tabularx}
\usepackage{comment}
\usepackage{hyperref}
\usepackage{enumitem}
\usepackage{soul}

\usepackage{ulem}

\hypersetup{
    colorlinks=true,
    linkcolor=red,
    citecolor=blue,
}

\begin{document}
\title{Fixing the dynamical evolution of self-interacting vector fields}
\author{Marcelo Rubio${}^{1,2}$}\email{mrubio@sissa.it}
\author{Guillermo Lara${}^{3}$}\email{glara@aei.mpg.de}
\author{Miguel Bezares${}^{4,5}$}\email{miguel.bezaresfigueroa@nottingham.ac.uk}
\author{Marco Crisostomi${}^{6,7}$}\email{mcrisost@caltech.edu}
\author{Enrico Barausse${}^{1,2}$}\email{barausse@sissa.it}
\affiliation{\vspace{0.2cm}
${}^{1}$SISSA, Via Bonomea 265, 34136 Trieste, Italy and INFN Sezione di Trieste\\
${}^{2}$IFPU - Institute for Fundamental Physics of the Universe, Via Beirut 2, 34014 Trieste, Italy
\\
${}^{3}$Max Planck Institute for Gravitational Physics (Albert Einstein Institute), Am M\"uhlenberg 1, 14476 Potsdam, Germany
\\
${}^{4}$Nottingham Centre of Gravity, University of Nottingham, University Park, Nottingham, NG7
2RD, UK\\
${}^{5}$School of Mathematical Sciences, University of Nottingham, University Park, Nottingham,
NG7 2RD, UK\\
${}^{6}$TAPIR, Division of Physics, Mathematics, and Astronomy, California Institute of Technology, Pasadena, CA 91125, USA\\
${}^{7}$Dipartimento di Fisica, Universit\`a di Pisa, Largo B. Pontecorvo 3, 56127 Pisa, Italy}

\begin{abstract}
Numerical simulations of the Cauchy problem for self-interacting massive vector fields often face instabilities and apparent pathologies. We explicitly demonstrate that these issues, previously reported in the literature, are actually due to the breakdown of the well-posedness of the initial-value problem. This is akin to shortcomings observed in scalar-tensor theories when derivative self-interactions are included. Building on previous work done for \(k\)-essence, we characterize the well-posedness breakdowns, differentiating between Tricomi and Keldysh-like behaviors. We show that these issues can be avoided by ``fixing the equations'', enabling stable numerical evolutions in spherical symmetry. Additionally, we show that for a class of vector self-interactions, no Tricomi-type breakdown takes place. Finally, we investigate initial configurations for the massive vector field which lead to gravitational collapse and the formation of black holes.
\end{abstract}

\maketitle


\section{Introduction}
\label{sec-Intro}

The detection of gravitational waves (GWs) from the merger of black holes (BHs) and neutron stars (NSs) \cite{LV_Catalog2019,LIGO_GW170817,LIGO_GW190521,LIGO_GW170817_2} has undoubtedly opened an exciting new window for testing fundamental physics in extreme conditions, and for understanding the nature of compact objects \cite{CardosoPani_LRR_2019}. Moreover, next-generation GW detectors \cite{LISA_WP_Baker:2019nia,LISA_Littenberg:2019grv,ET_Branchesi:2023mws} will be capable of probing modifications to General Relativity (GR) with higher precision, and will provide new data that may prove crucial for unveiling fundamental questions about the internal structure of NSs, the validity of the Kerr hypothesis~\cite{Israel:1967wq, Hawking:1971vc, Robinson:1975bv}
and the possible existence of  ``exotic'' compact objects, many of which have been proposed in the literature \cite{Bezares:2024btu,Berti2006,ECO1,ECO2,ECO3,ECO4,Kesden2005}. 

An example of the latter are boson stars (BSs) \cite{JETZER1992163,BSs1,Schunck:2003kk,PhysRevD.80.084023,PhysRevLett.57.2485}, defined as regular gravitating solitons \cite{PhysRev.172.1331}. 
While BSs often refer to spin-0 scalar Bose-Einstein condensates~\cite{PhysRevD.38.2376,PhysRevD.39.1257,lee1992nontopological,Lee:2008jp,Lee:2008ab}, spin-1 vector BSs, commonly referred to as \textit{Proca stars} (PSs), have also been put forward as models of ``black-hole mimickers'' \cite{BRITO2016291,Herdeiro:2023lze,PhysRevLett.123.221101,Herdeiro:2023wqf}. In fact, shadows and lensing around PSs have been simulated and are compatible with current constraints from the Event Horizon Telescope (EHT) \cite{Pheno_Sengo2022,Sengo:2024pwk}.
In addition, dynamical mechanisms for the formation of PSs via gravitational cooling have also been put forward in the literature  \cite{Pheno2_Herdeiro2023,PhysRevD.98.064044,Shen:2016acv}.

The simplest models for compact exotic stars are given by a complex scalar or vector field minimally coupled to gravity, and they lead to peculiar stability properties. For instance, scalar BSs are generally stable under non-spherical perturbations, but static, spherical PSs might display generic instabilities \cite{Pheno2_Herdeiro2023}. Although some of these issues can be resolved for stationary rotating stars \cite{PhysRevLett.123.221101}, the situation notably changes  when accounting for self-interactions. This fact has motivated deeper investigations on more general Einstein-Proca systems (see for instance \cite{Herdeiro:2023lze}). Additionally, explorations on self-interacting massive vector fields reported superradiant instabilities due to exponential amplification of bound state modes, and energy/angular momentum extraction from spinning BHs \cite{EastPRL17,East17,Dolan18}.

A numerical study of self-interacting Proca fields on a Kerr background was performed in \cite{Proca_CloughPRL2022}, also in the context of superradiance, but reporting instabilities which were attributed to the development of ghost (or tachyonic) modes. Similar issues were found afterwards, e.g. in \cite{Proca_Coates1_PRL2022,Barvinsky:1989wy,Zong-Gang22}. In Ref.~\cite{ShortPaper_Barausse2022}, some of us pointed out that these pathologies are actually due not to ghost/tachyonic modes but to the breakdown of the well-posedness of the corresponding Cauchy problem. 
Indeed, although the principal part of the equations of motion governing massive vector theories might at first seem the same as for Maxwell equations (and therefore innocuous), self-interactions drastically modify it and potentially make the Cauchy problem ill-posed. This was confirmed by further studies~\cite{Proca_Coates2_PRL2023, Proca_Coates3_PRD2023, Proca_Coates4_PRD2023,Proca_Silva2022,Aoki22}, which report  instabilities occurring when the dynamical fields reach a point where the initial-value problem is mathematically no longer well-defined.

Another key point raised by Ref.~\cite{ShortPaper_Barausse2022} is that the well-posedness of the Cauchy Problem of self-interacting massive vector fields is completely analogous to that of \textit{k}-essence theories (i.e., scalar theories with first-order derivative self-interactions), which was thoroughly studied and characterized in the last few years \cite{Bernard19,Bezares_Kdynamics2020,Lotte_PRL21,Bezares21KS,Bezares:2021dma,Figueras:2020dzx}. Ref.~\cite{ShortPaper_Barausse2022}  showed that this analogy becomes evident when restoring the $U(1)$ gauge symmetry via a St\"uckelberg transformation.
Indeed,  this transformation
highlights  
that vector field self-interactions actually hide \textit{derivative} self-interactions of the St\"uckelberg scalar field (which corresponds to the longitudinal mode of the vector field), radically modifying the principal part of the evolution system.
As a result,  strategies similar to those successfully adopted for \textit{k}-essence could also be implemented for simulating self-interacting vector fields, avoiding the aforementioned breakdown of the initial-value problem.  

In this work, and following \cite{ShortPaper_Barausse2022}, we carry out numerical investigations of massive vector fields in the St\"uckelberg formulation, and study the well-posedness of the associated initial-value problem. In particular, we show an explicit example of a Tricomi-type evolution (i.e., one leading to a change of character of the equations from hyperbolic to parabolic or elliptic) and two ways of avoiding it. The first one is provided by the ``fixing-the-equations'' technique \cite{Fix_CayusoPRD2017}, which consists of modifying the principal part of the evolution system by introducing extra dynamical fields satisfying \textit{ad hoc} differential equations. This approach has already been applied with success to several gravitational theories \cite{Fix_Franchini2022,Bezares21KS,CayusoNonIterative20,MemoUV2022,Lara:2024rwa,Corman:2024cdr}. It was also considered in Ref. \cite{PhysRevD.108.L101501},  in the context of massive fields with quadratic self-interactions, performing evolutions in flat space and in one spatial dimension. The second alternative explored in the present work is to account for a cubic self-interaction in the Lagrangian, without having to ``fix'' the equations. In both cases, we perform full numerical relativity simulations in spherical symmetry. We also explore configurations leading to gravitational collapse and the formation of black holes in theories with cubic self-interactions.

The outline of the paper is the following: in section \ref{sec-Preliminaries}, we revisit the theory of self-interacting massive vector fields in the St\"uckelberg language,  its corresponding Cauchy problem, and the ``fixing-the-equations'' technique. In section \ref{sec-Numerical-Implementation}, we describe our numerical set-up, initial data and boundary conditions in detail. The main results of the paper are reported in section \ref{sec-Results}, which is divided into three parts. Firstly, we show numerically an example of a Tricomi-type breakdown of the Cauchy problem and present how the ``fixing'' approach restores the well-posedness during the evolution. Secondly, we allow for a cubic self-interaction in the Lagrangian, and show that no Tricomi-type breakdown is developed during the whole evolution. Finally, we study gravitational collapse in the case of cubic self-interactions, reporting an explicit example of an initial configuration whose evolution leads to the formation of a black hole. An overall discussion of our results is left for Section \ref{sec-Conclusions}. Throughout this work, we use the mostly-plus signature for the spacetime metric, $(-,+,+,+)$, while the employed units are specified at the beginning of Section \ref{sec-Results}.

\section{Preliminaries}
\label{sec-Preliminaries}

\subsection{Massive vector fields}

We study the dynamics of a self-interacting real vector field $A^{\mu}$ with mass $m$, coupled to GR. The corresponding action is given by
\begin{eqnarray}\label{action-proca}
    S &=& \int d^4x \sqrt{-g}\left[\frac{R}{16\pi G} -\frac{1}{4}F_{\mu\nu}F^{\mu\nu} \right. \\
    && \left. \qquad\qquad\qquad\qquad -\frac{m^2}{2}A_{\mu}A^{\mu} + V(A_{\mu}A^{\mu})\right] \,, \nonumber
\end{eqnarray}
where $F_{\mu\nu}:=\nabla_{\mu}A_{\nu}-\nabla_{\nu}A_{\mu}$ is the vector strength, 
and the self-interaction potential is
\begin{equation}\label{potential-self-int}
    V(A_{\mu}A^{\mu}) = \frac{\beta}{4} (A_{\mu}A^{\mu})^2 - \frac{\gamma}{8\Lambda^2} (A_{\mu}A^{\mu})^3.
\end{equation}
Here, $\Lambda$ is the energy scale suppressing the higher-order interactions, whereas 
$\beta$ and $\gamma$ are $\mathcal{O}(1)$ dimensionless coupling constants for the quadratic and cubic interactions, respectively. The field $A_\mu$ propagates two transverse and one longitudinal modes. Although action (\ref{action-proca}) lacks the $U(1)$ gauge symmetry $A_\mu \to A_\mu + \nabla_\mu f$ of Maxwell electrodynamics, it is possible to restore it at the cost of introducing an extra dynamical field. In fact, by performing the St\"uckelberg transformation
\begin{equation}\label{stueckelberg-transf}
    A_{\mu} \to A_{\mu} - \frac{1}{m}\nabla_\mu\phi,
\end{equation}
the product $A_\mu A^\mu$ transforms as
\begin{equation}
    A_\mu A^\mu \to A_\mu A^\mu - \frac{2}{m}A^\mu\nabla_\mu \phi + \frac{1}{m^2} \nabla_\mu \phi \nabla^\mu \phi ,
\end{equation}
and this particular mixing between $A_\mu$ and $\phi$ remains invariant under the  gauge transformation
\begin{equation}\label{gauge-transf-A-phi}
    \begin{pmatrix}
    A_\mu \\
    \phi 
    \end{pmatrix}
    \to
    \begin{pmatrix}
    A_\mu + \nabla_\mu f \\
    \phi + m f
    \end{pmatrix}
    .
\end{equation}
Thus, the theory obtained by replacing (\ref{stueckelberg-transf}) into the action (\ref{action-proca}) results invariant under (\ref{gauge-transf-A-phi}). This symmetry induces a gauge freedom in the choice of the dynamical fields, which can be fixed in different ways. By adopting for instance the \textit{unitary gauge} (i.e., $\phi=0$), one would recover the original Lagrangian. The \textit{Lorenz gauge} (i.e., $\nabla_\mu A^\mu=0$) is instead useful for decoupling the vector and scalar degrees of freedom at high energies.

It is worthwhile to introduce the gauge invariant derivative of the scalar field
\begin{equation}\label{derivative-operator-stueckelberg}
    D_\mu\phi := \nabla_\mu \phi - mA_\mu \,,
\end{equation}
so that, under the transformation (\ref{stueckelberg-transf}), one has
\begin{equation}
    m^2A_\mu A^\mu \to D_\mu\phi D^\mu \phi=:X \,.
\end{equation}
Then, after the St\"uckelberg's procedure, the action (\ref{action-proca}) can be re-written as
\begin{equation}\label{action-stueckelberg}
    S = \int{d^4x \sqrt{-g}\left[\frac{R}{16\pi G} -\frac{1}{4}F_{\mu\nu}F^{\mu\nu} + K(X)\right]} \,,
\end{equation}
where
\begin{equation}\label{lagrangian-stueckelberg-K(X)}
     K(X) := -\frac{1}{2}X + \frac{\beta}{4 m^4}X^2 - \frac{\gamma}{8\Lambda^2 m^6}X^3 \,.
\end{equation}
Variations of (\ref{action-stueckelberg}) with respect to the dynamical fields $\{ g_{\mu\nu}, A^\mu, \phi\}$ lead to the equations of motion
\begin{eqnarray} \label{full-eqs}
    && G_{\mu\nu} = 8\pi G \left(T^{\text{\tiny{(EM)}}}_{\mu\nu} + T^{(\phi)}_{\mu\nu}\right) \,, \nonumber \\
    && \nabla_\mu F^{\mu\nu} = J_{(\phi)}^\nu \,, \\
    && \nabla_\mu\left[K'(X)D^\mu\phi\right] = 0\,, \nonumber 
\end{eqnarray}
where $G_{\mu\nu}$ is the Einstein's tensor,
\begin{equation}
    T^{\text{\tiny{(EM)}}}_{\mu\nu} = F_{\mu\rho}F_{\nu}{}^\rho - \frac{1}{4}F_{\rho\sigma}F^{\rho\sigma}g_{\mu\nu}
\end{equation}
is the energy-momentum tensor associated to $F_{\mu\nu}$, and
\begin{equation}\label{Tab-scalar-field}
    T^{(\phi)}_{\mu\nu} = K(X)g_{\mu\nu} - 2K'(X)D_\mu\phi D_\nu\phi
\end{equation}
is the one associated with the St\"uckelberg field. The source of the equation for $F^{\mu\nu}$ is
\begin{equation}
    J_{(\phi)}^\mu = -2mK'(X)D^\mu \phi \,,
\end{equation}
where a prime denotes a derivative with respect to $X$.

Notice that, because $D_\mu\phi$ in Eq.~(\ref{derivative-operator-stueckelberg}) contains the vector field, the action (\ref{action-stueckelberg}) still retains the full mixing between $\phi$ and $A^\mu$. However, written in this form, it becomes easy to single out the respective principal parts, which correspond to the high-energy/momentum  decoupling limit. Because this limit  determines the well-posedness of the corresponding initial-value problem, one can see from the action (\ref{action-stueckelberg}) that the scalar sector exhibits first-order \textit{derivative} self-interactions (as in \(k\)-essence theories), which may lead to pathologies during dynamical evolutions \cite{Bernard19,Bezares_Kdynamics2020,Bezares21KS,MemoUV2022,ShortPaper_Barausse2022,Lotte_PRL21}. This calls for a full well-posedness analysis of the equations (\ref{full-eqs}), particularly the scalar sector.

\subsection{The Cauchy problem of \(k\)-essence theories}

We will now briefly review the initial-value problem for \(k\)-essence theories, following Refs.~\cite{Bernard19,Bezares_Kdynamics2020,Bezares21KS,Bezares21KS,Figueras:2020dzx}.

Generally, we refer to the Cauchy problem of a system of partial differential equations as \textit{well-posed} \cite{Hadamard1908} if, for any given initial-data set (e.g. at $t=t_0$), there exists a time interval $I=[t_0,T]$ in which: \textit{i)} a solution exists; \textit{ii)} it is unique; \textit{iii)} it is a continuous function of the initial data (with respect to the topologies where the data and solutions are defined). The evolution is governed by the \textit{principal part} of the system of equations, as it contains all the information about the propagation speeds of the different modes \cite{Kreiss70,geroch1996partial}. An algebraic strategy to assess these non-trivial mathematical conditions is provided by the concept of \textit{hyperbolicity} \cite{Friedrichs54,Kreiss70,friedrichs1971systems,geroch1996partial}; that is, a set of algebraic conditions that the principal part must satisfy for the system to admit a well-posed (sometimes said \textit{hyperbolic}) initial-value formulation. In the particular case of system (\ref{full-eqs}), the scalar sector governs the hyperbolicity of the system, and might be the cause of a possible change of character during the evolution. In fact, its principal part can be recast as a modified Klein-Gordon equation, with an effective metric $\gamma^{\mu\nu}$ given by
\begin{equation}
    \gamma^{\mu\nu}=g^{\mu\nu} + \frac{2K''(X)}{K'(X)}D^\mu\phi D^\nu\phi \,.
\end{equation}
Thus, the system is strongly hyperbolic if and only if $\det(\gamma^{\mu\nu})<0.$ Nevertheless, nothing prevents this condition from potentially failing during the evolution, even when starting from regular initial data sets. The highly non-linear evolution could cause the determinant of the effective metric to become zero (implying that the equations become parabolic), or it could give rise to very large (or even diverging) characteristic speeds. The breakdown of the Cauchy problem due to these shortcomings are usually referred to as \textit{Tricomi} and \textit{Keldysh} types, respectively \cite{Bernard19,Bezares_Kdynamics2020,Ripley19,Ripley:2022cdh}. 

Let us analyze under which conditions these pathologies could appear, in the particular case of a spherically symmetric configuration. The corresponding line element can be expressed as
\begin{equation}\label{metric-sphsym}
    ds^2 = -\alpha(t,r)^2 dt^2 + g_{rr}(t,r) dr^2 + r^2g_{\theta\theta}(t,r)d\Omega^2,
\end{equation}
where $\alpha$ is the lapse function, $g_{rr}$ and $g_{\theta\theta}$ are positive fields, and $d\Omega^2$ is the line element of the unit sphere. As pointed out in Ref. \cite{Bezares_Kdynamics2020}, the determinant of the effective metric reads
\begin{equation}\label{det-gamma}
    \det(\gamma^{\mu\nu}) = - \frac{1}{\alpha^2g_{rr}}\left(1+\frac{2K''(X)}{K'(X)}X\right),
\end{equation}
and its dynamical evolution can be assessed by examining the eigenvalues of $\gamma^{\mu\nu}$, namely
\begin{equation}\label{k-essence-eigenvalues}
    \lambda_\pm = \frac{1}{2}\left[\gamma^{tt}+\gamma^{rr}\pm\sqrt{(\gamma^{tt}-\gamma^{rr})^2+4(\gamma^{tr})^2}\right] \,.
\end{equation}
The characteristic speeds of the scalar equation of system (\ref{full-eqs}) are given by
\begin{equation}
    V_\pm = -\frac{\gamma^{tr}}{\gamma^{tt}}\pm\sqrt{\frac{-\det(\gamma^{\mu\nu})}{(\gamma^{tt})^2}} \,. \label{Vpm}
\end{equation}
The hyperbolicity condition $\det(\gamma^{\mu\nu})<0$ restricts the possible values of the coupling constants for the function $K(X)$ given in Eq. (\ref{lagrangian-stueckelberg-K(X)}). As an example, looking at Eq. (\ref{det-gamma}) one can see that the choice $\beta>0$ and $\gamma=0$ could give rise to the Tricomi pathology for certain initial data.
On the other hand, choosing for instance $\beta=0$ and $\gamma > 0$ avoids it altogether, although a Keldysh type behavior (i.e., $\gamma^{tt} \to 0$ in Eq. (\ref{Vpm})) can still occur. Nonetheless, we stress that the diverging  speeds 
characterizing the Keldysh type behavior do not violate hyperbolicity. The system remains hyperbolic, but in practice numerical evolutions become impossible as the Courant-Friedrichs-Lewy (CFL) condition forces the time step to vanish if the evolution
is performed with an explicit method. We also notice
that Keldysh type behaviors depend on gauge; i.e., they can be (in principle) avoided with a judicious gauge choice~\cite{Bezares_Kdynamics2020,Bezares:2021dma}. In this sense, Keldysh type breakdown of well-posedness could be a practical problem but not a fundamental pathology. 
In this work, we numerically explore how these issues appear in the context of self-interacting massive vector fields and show possible ways to cure them.

\subsection{The \textit{fixing-the-equations} approach}

An operational method to deal with the ``ill-suited’’ evolutions reviewed before (which generally appear in the context of modified theories of gravity) is through the so-called \textit{fixing-the-equations} approach \cite{Fix_CayusoPRD2017,Fix_Allwright2019,Fix_Cayuso2_PRD2020}. This possibility has been proven to be useful in the strong, non-linear, and highly dynamical regime of $k-$essence theories \cite{Bezares21KS,MemoUV2022}.

Inspired by relativistic theories for dissipative fluids \cite{Israel761,Israel762,Israel79} and by numerical methods to deal with interfaces between touching numerical grids \cite{LuisPC}, the idea of the fixing-the-equations approach is to modify \textit{ad hoc} the evolution equations (with particular focus on their higher-derivative contributions), so that the high-frequency modes are controlled. In this way, the norm of the solution remains finite and bounded at all times, rendering the system well-posed. To do so, one introduces new auxiliary variables and a timescale on which they settle to their true/physical values through a driver equation. This method allows for capturing the main features of the non-linear behavior of the system and can be easily implemented numerically.

In \cite{MemoUV2022}, this method has been utilized for simulating spherical collapse in quadratic \textit{k}-essence after a Tricomi-type breakdown occurs. In \cite{PhysRevD.108.L101501}, it was also implemented for evolving massive vector fields with quartic self-interactions in one spatial dimension in flat space. With a similar spirit, here we introduce a variable $\Sigma$, a timescale $\tau$ and a driver equation in order to replace the evolution for the St\"uckelberg field in (\ref{full-eqs}) by the system
\begin{eqnarray}
    \nabla_\mu\left(\Sigma D^\mu\phi\right)&=&0\,, \label{fixed-scalar} \\
    \partial_t\Sigma + \frac{\Sigma-K'(X)}{\tau} &=& 0.
\end{eqnarray}
This modification damps all the modes with frequencies larger than $1/\tau$ and forces $\Sigma\to K'(X)$ as $\tau\to 0$, resulting in a strongly hyperbolic evolution. In fact, for the spherically symmetric Ansatz (\ref{metric-sphsym}), and for $\Sigma\neq 0$, the system admits the characteristic speeds
\begin{equation}
    V^o_{\scriptsize{\mbox{fixed}}}=0\,, \quad V^\pm_{\scriptsize{\mbox{fixed}}}=\pm\frac{\alpha}{\sqrt{g_{rr}}}\,,
\end{equation}
which agree with the ones of the (spherically-symmetric) Klein-Gordon equation in GR.

It is worthwhile to mention at this point that a different approach for resolving the Cauchy problem of self-interacting massive vector fields is provided -- for the cases where it is known -- by a well-posed ultraviolet completion (UV) of the low-energy effective field theory (EFT). Indeed, for certain self-interactions, this is given by the Abelian Higgs model, and the corresponding dynamical evolution has been explored numerically in \cite{Higgs_East:2022ppo,Corman:2024cdr,PhysRevD.108.L101501}. The comparison of the evolutions given by the low-energy fixed EFT and the UV Higgs one is left for future investigation.


\section{Numerical Implementation}
\label{sec-Numerical-Implementation}

In this section, we give details about the methodology used for the numerical simulations, including the evolution equations in spherical symmetry, the initial data and the implemented boundary conditions. 

We performed evolutions of both the original (\ref{full-eqs}) and the fixed (\ref{fixed-scalar}) systems introduced in the previous section. For doing so, we extended our previous code used in \cite{Bezares_Kdynamics2020,Bezares21KS,Bezares21KS,MemoUV2022} by including the equations for the electromagnetic sector. This code implements a high-resolution shock-capturing (HRSC) finite-difference scheme, originally developed in \cite{Alic:2007ev}. It was used for simulating black holes and for studying the dynamics of boson stars, fermion-boson stars and anisotropic stars  \cite{Z3_Bernal09,Z3_Valdez-Alvarado2013,Raposo:2018rjn}. The numerical scheme can be considered as a fourth-order finite-difference one plus a third-order adaptive dissipation, with the dissipation coefficient given by the maximum propagation speed at each grid point. Time evolution is performed using the method of lines, with a third-order Runge-Kutta scheme.
In the context of this paper, the choice of an HRSC method is motivated by the fact that $k-$essence theories might develop caustics/shocks during evolution, even from smooth initial data, as reported in \cite{Felder:2002sv,Reall:2014sla,Babichev:2017lrx,Tanahashi:2017kgn}.

\subsection{Evolution equations}

In order to set up evolution equations, we perform a $(3+1)$-decomposition by introducing a foliation $\{\Sigma_t\}_{t\in\mathbb{R}}$ of spatial hypersurfaces with normal $n_{\mu} = (-\alpha,\,0)$. We define the extrinsic curvature of each $\Sigma_t$ as $K_{ij} = -\mathcal{L}_n h_{ij}/2$, where $h_{ij}$ is the induced metric. 

By taking the spherically-symmetric Ansatz (\ref{metric-sphsym}) for the spacetime metric, the trace of $K_{ij}$ reads
\begin{equation}
	K = K_r{}^r + 2K_\theta{}^\theta.
\end{equation}
We denote the parallel and orthogonal projections of $A_\mu$ (with respect to $n^\mu$) as
\begin{eqnarray} \label{A-parallel-perp}
    A_\parallel &=& -n^\mu A_\mu\,, \nonumber \\
    A_\perp &=& \delta_r{}^\mu A_\mu\,,
\end{eqnarray}
respectively.

\subsubsection{Metric evolution}

All the simulations were performed using the Z3 formulation \cite{PhysRevD.66.084013}, which is a strongly-hyperbolic first-order reduction for Einstein's equations. It can be obtained from the Z4 formulation by a \textit{symmetry breaking} procedure.
We made use of the spherically-symmetry version developed in \cite{Z3_Bernal09,Z3_Valdez-Alvarado2013}, and written in flux-conservative form \cite{PhysRevD.40.1022}. 
The dynamical variables for the metric evolution are $\{\alpha_r,D_{rr}{}^r,D_{r\theta}{}^{\theta},Z_r,K_r{}^r,K_\theta{}^\theta\}$, where $\alpha_r=\partial_r\alpha/\alpha$,
\begin{equation} \label{first-order-vars}
    D_{ri}{}^i = \frac{g^{ii}}{2}\partial_r g_{ii}, 
\end{equation}
and $Z_r$ is the radial component of $Z_i$\footnote{We stress that the Z3 formulation considers and additional evolution field, $Z_i$, which accounts for the Momentum constraint, and thus it should remain close to zero throughout the whole numerical evolution. In spherical symmetry, only the radial component $Z_r$ is dynamical. We refer the reader to Refs. \cite{Z3_Bernal09,Z3_Valdez-Alvarado2013}, where the explicit set of field equations in this formulation is displayed.}. Finally, we fix the gauge freedom by setting a ``$1+\log$'' slicing condition \cite{PhysRevLett.75.600} in normal coordinates (with zero shift).

\subsubsection{Scalar sector}

We do a first-order reduction of the scalar field equation, by introducing the variables
\begin{equation}
    \Pi = - \dfrac{\partial_t \phi}{\alpha}\,, \quad \Phi = \partial_r \phi.
\end{equation}
It is also useful to define the auxiliary ``shifted'' variables
\begin{eqnarray}
    \tilde{\Pi} &\equiv& \Pi - m A_\parallel\\
    \tilde{\Phi} &\equiv& \Phi - m A_\perp, 
\end{eqnarray}
as they allow to express the kinetic term as
\begin{equation}
    X = -\tilde{\Pi}^2 + g^{rr} \tilde{\Phi}^2\,.    
\end{equation}

In order to deal with shocks, we write the evolution equations in flux-conservative form,
\begin{equation}
    \partial_t \bm{U}_{(\phi)} + \partial_r \bm{F}_{\mathrm{(\phi)}}(\bm{U}_{\mathrm{(\phi)}}) + \bm{S}_{\mathrm{(\phi)}}(\bm{U}_{\mathrm{(\phi)}})=\bm{0},
\end{equation}
where
\begin{equation}
    \bm{U}_{(\phi)} =
    \begin{bmatrix}
    \phi \\
    \Phi \\
    \sqrt{g_{rr}} g_{\theta \theta} K'(X)\tilde{\Pi}
    \end{bmatrix}
    \,,
\end{equation}
\begin{equation}
    \bm{F}_{\mathrm{(\phi)}}(\bm{U}_{\mathrm{(\phi)}}) =
    \begin{bmatrix}
    0 \\
    \alpha \Pi \\
   \displaystyle\frac{\alpha g_{\theta\theta}}{\sqrt{g_{rr}}} K'(X)\tilde{\Phi}
    \end{bmatrix}
    \,,
\end{equation}
and
\begin{equation}
    \bm{S}_{(\phi)}(\bm{U}_{(\phi)}) =
    \begin{bmatrix}
     \alpha \Pi\\
    0 \\
    \displaystyle\frac{2}{r}\frac{\alpha g_{\theta\theta}}{\sqrt{g_{rr}}} K'(X)\tilde{\Phi}
    \end{bmatrix}.
\end{equation}

Finally, the projections of the energy-momentum tensor (\ref{Tab-scalar-field}) associated to the scalar field are
\begin{align}
    \tau^{(\phi)} & = - K(X)- 2 K'(X) \tilde{\Pi}^2, \\
    {S^{(\phi)}}^{r} 
    &= -2 g^{rr} K'(X) \tilde{\Pi} \tilde{\Phi}, \\
    {{S^{(\phi)}}_{r}}^{r} & = K(X) - 2 g^{rr} K'(X) \tilde{\Phi}^2, \\  
    {{S^{(\phi)}}_{\theta}}^{\theta} & = K(X).
\end{align}

\subsubsection{U(1) vector sector}

The equations for the electromagnetic sector are usually given in terms of the electric and magnetic fields $E^\mu = -t^\nu F_\nu{}^\mu$ and $B^\mu = -t^\nu {}^*F_\nu{}^\mu$, respectively. These are measured with respect to an Eulerian observer determined by a timelike, unitary and future-pointing frame $t^{\mu}$. It  is however natural to consider the evolution of $A_\mu$, such that $F_{\mu\nu} = 2\nabla_{[\mu}A_{\nu]}$. In spherical symmetry, the only non-trivial components are $A_\parallel$ and $A_\perp$, introduced in (\ref{A-parallel-perp}). This choice automatically implies that $B^i = \varepsilon^{ijk}D_jA_k = 0$, and the only nontrivial component for the electric field is the radial one, namely $E^r$. Moreover, since the scalar equation in system (\ref{full-eqs}) is coupled with the vector field, we also need to provide evolution equations for it. Like we did for the scalar sector, we express them as
\begin{equation}
    \partial_t\, \bm{U}_{\mathrm{E}} + \partial_r\, \bm{F}_{\mathrm{E}}(\bm{U}_{\mathrm{E}}) + \bm{S}_{\mathrm{E}}(\bm{U}_{\mathrm{E}})=\bm{0},
\end{equation}
for the variables
\begin{equation}
    \bm{U}_{\mathrm{E}} =
    \begin{bmatrix}
    E^r \\
    A_\parallel \\
    A_\perp
    \end{bmatrix},
\end{equation}
with radial flux
\begin{equation}\label{EM-fluxes}
    \bm{F}_{\mathrm{E}}(\bm{U}_{\mathrm{E}}) =
    \begin{bmatrix}
    0 \\
    \displaystyle \frac{\alpha}{g_{rr}}A_\perp \\
    \alpha A_\parallel
    \end{bmatrix},
\end{equation}
and sources
\begin{equation}
    \bm{S}_{\mathrm{E}}(\bm{U}_{\mathrm{E}}) =
    \begin{bmatrix}
    \alpha (4\pi j_e - K E^r) \\
     \dfrac{\alpha A_\perp}{g_{rr}}\left(\dfrac{2}{r} + 2 {D_{r\theta}}^{\theta} + {D_{rr}}^r\right)-\alpha K A_\parallel\\
    \alpha\,g_{rr}\,E^r
    \end{bmatrix}.
\end{equation}

The Gauss constraint $D_iE^i = 4\pi\rho_e$ reads
\begin{equation}\label{gauss}
    \partial_r E^r + \left(\dfrac{2}{r} + 2 {D_{r \theta}}^{\theta} + {D_{rr}}^{r}\right)E^r - 4 \pi \rho_e = 0,
\end{equation}
and the electromagnetic sources are explicitly given by
\begin{eqnarray}
    4 \pi \rho_e &=& - 2 m K'(X) \tilde{\Pi}, \label{EM-sources} \\
    4 \pi j_e &=& - 2 m K'(X)g^{rr}\tilde{\Phi}.
\end{eqnarray}

\subsection{Boundary conditions}

We consider approximate outgoing Sommerfeld conditions for the outer boundary of our numerical domain, for which a detailed analysis of the characteristic structure of the system is required. The electromagnetic sector admits the eigenfields $\{E^r,\,A_\pm\}$, where
\begin{equation}\label{eigenfields}
    A_\pm = \frac{1}{2}\left(A_\perp \pm \sqrt{g_{rr}}A_\parallel\right),
\end{equation}
and with respective eigenvalues $\{0,\,\pm \alpha/\sqrt{g_{rr}}\}$.
Inverting Eq. (\ref{eigenfields}), we get
\begin{eqnarray}
    A_\parallel &=& \frac{A_{+} - A_{-}}{\sqrt{g_{rr}}},\\
    A_\perp &=& A_{+} + A_{-}.    
\end{eqnarray}

Following a similar idea implemented in \cite{Torres:2014fga}, we 
apply outgoing boundary conditions by requiring
\begin{align}
    \partial_t A_{-} +  v_{-} \partial_r A_{-} + \dfrac{v_{-}}{r} A_{-} = 0,
\end{align}
where \(v_{-} = - \alpha / \sqrt{g_{rr}}\), and rewriting it in flux-conservative form. For the rest of the system, we impose maximally dissipative boundary conditions, for which all incoming fields at the outer boundary are suppressed,  thus damping spurious reflections as much as possible.

\subsection{Initial data}

For the initial-data set, we need to prescribe values for all the dynamical fields in such a way that  the Hamiltonian, momentum and Gauss constraints are satisfied at $t=0$. For the metric fields, we set $\alpha(t=0,r)=1$, we write the spatial metric in isotropic coordinates, yielding the line element $dh^2 = \psi^4(r)(dr^2 + r^2 d\Omega^2)$, and solve for the conformal factor $\psi(r)$ from the Hamiltonian constraint. 

We consider a stationary initial configuration for which $K_r{}^r=K_\theta{}^\theta=\Pi=0$, and set up the scalar field in two ways:
\begin{itemize}
    \item \textbf{ID Type I:} we consider a pulse in $\Phi=\partial_r\phi$ given by 
        \begin{equation} \label{ID-typeI}
            \Phi(t=0,r)=A\,\exp\left[-\frac{(r-r_c)^2}{\sigma^2}\right]\cos\left(\frac{\pi}{10}r\right);
        \end{equation}
    \item \textbf{ID Type II:} we set the pulse in the scalar field itself, with the form
        \begin{equation} \label{ID-typeII}
            \phi(t=0,r)=A\,\exp\left[-\frac{(r-r_c)^2}{\sigma^2}\right]\sin\left(\frac{r-r_c}{\sqrt{2}\,\sigma}\right),
        \end{equation}
    and thus $\Phi(t=0,r)=\partial_r\phi|_{t=0}$.
\end{itemize}

While the first configuration was used in \cite{MemoUV2022} to induce a Tricomi-type breakdown in quadratic \textit{k}-essence, the second one was implemented in \cite{Bezares_Kdynamics2020} for simulating gravitational collapse when cubic derivative self-interactions are taken into account.

For the electromagnetic variables, we choose the simplest possible initial configuration, given by $E^r=A_\parallel=A_\perp=0$, for which the Gauss constraint is automatically satisfied, as $\rho_e = 0$ (see Eq. (\ref{EM-sources})). With this choice for the initial fields, the momentum constraint is trivially satisfied, while the Hamiltonian constraint yields a second-order elliptic equation for the conformal factor $\psi(r)$, given by
\begin{equation}\label{HamConstraint}
    \frac{1}{r^2}\frac{\partial}{\partial r}\left(r^2\frac{\partial\psi}{\partial r}\right) - 2\pi G K(X)\psi^5=0.
\end{equation}
We solve this equation by shooting in $\psi(r=0)$, requiring regularity at the origin and imposing a Robin boundary condition at infinity; i.e.,
\begin{equation}
    \lim_{r\to 0}{\frac{\partial\psi}{\partial r}}=0\,, \quad
    \lim_{r\to \infty}{\left(\psi + r\frac{\partial\psi}{\partial r}\right)}=1\,.
\end{equation}

Finally, for the ``fixed'' evolution, we set the initial value of the auxiliary field $\Sigma$ to
\begin{equation}
    \Sigma(t=0,r)=K'(X)|_{t=0}\,.
\end{equation}


\begin{figure*}
  \centering
  \includegraphics[scale=0.6]{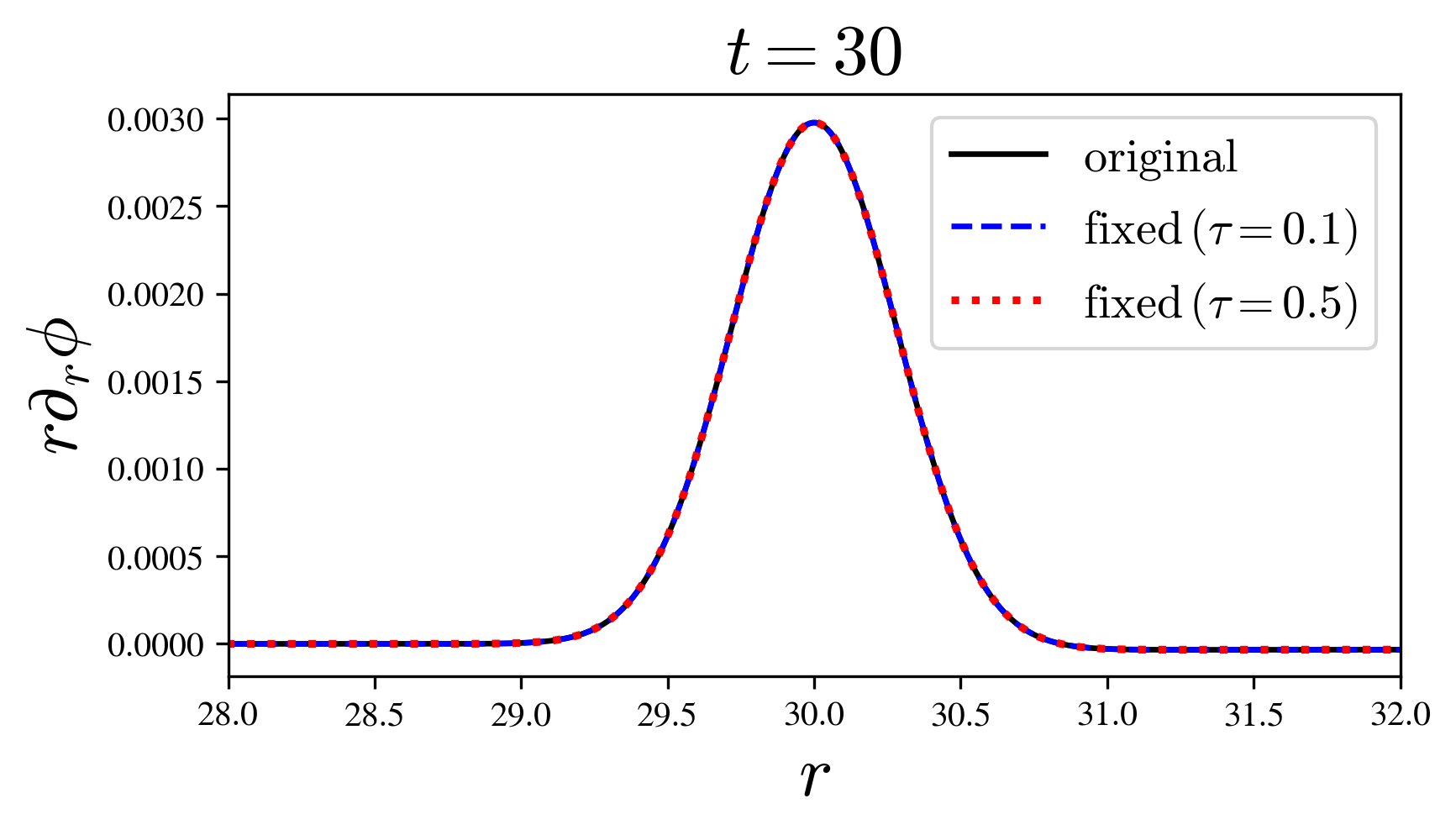}
  \includegraphics[scale=0.6]{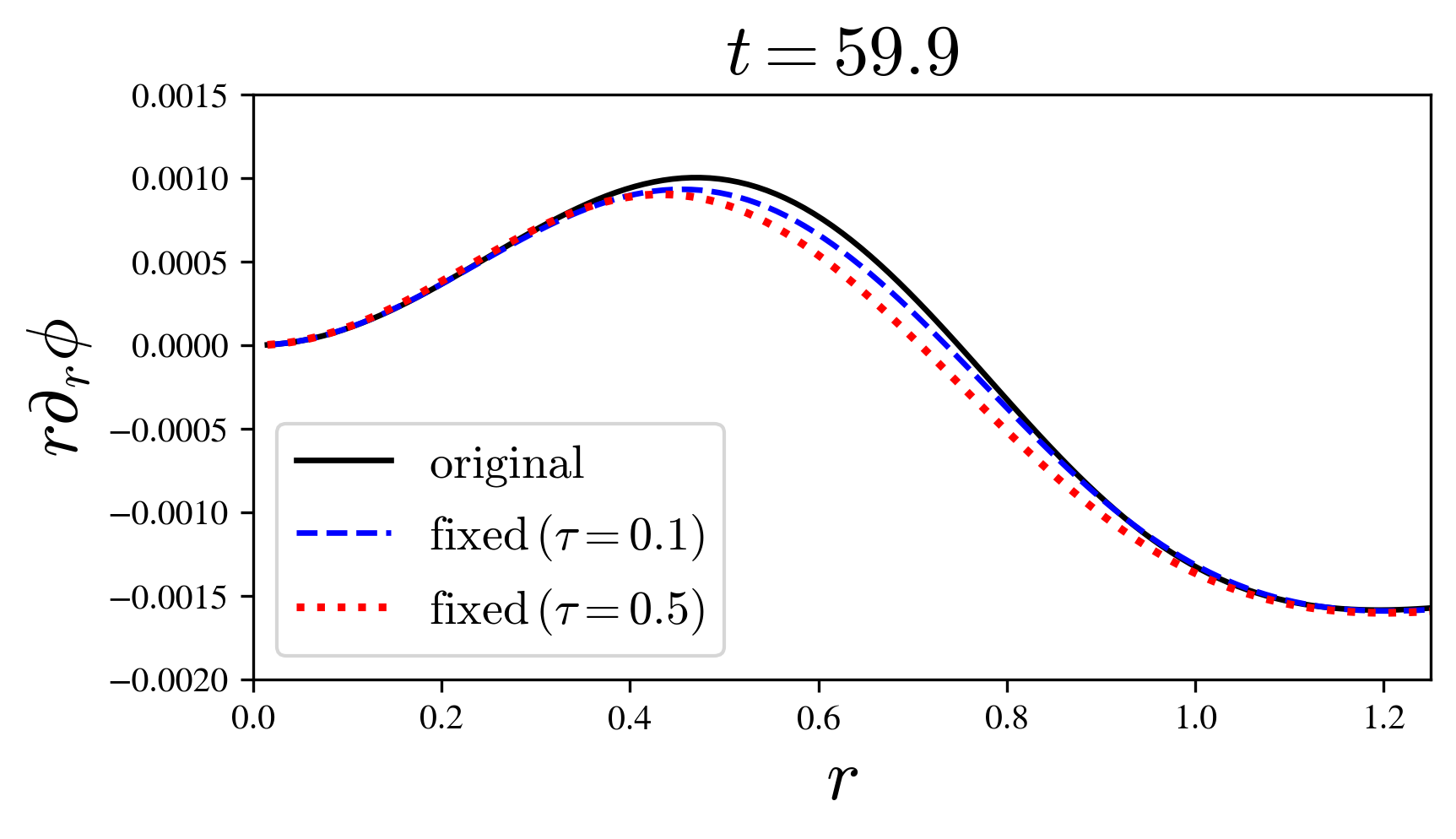}
  \includegraphics[scale=0.6]{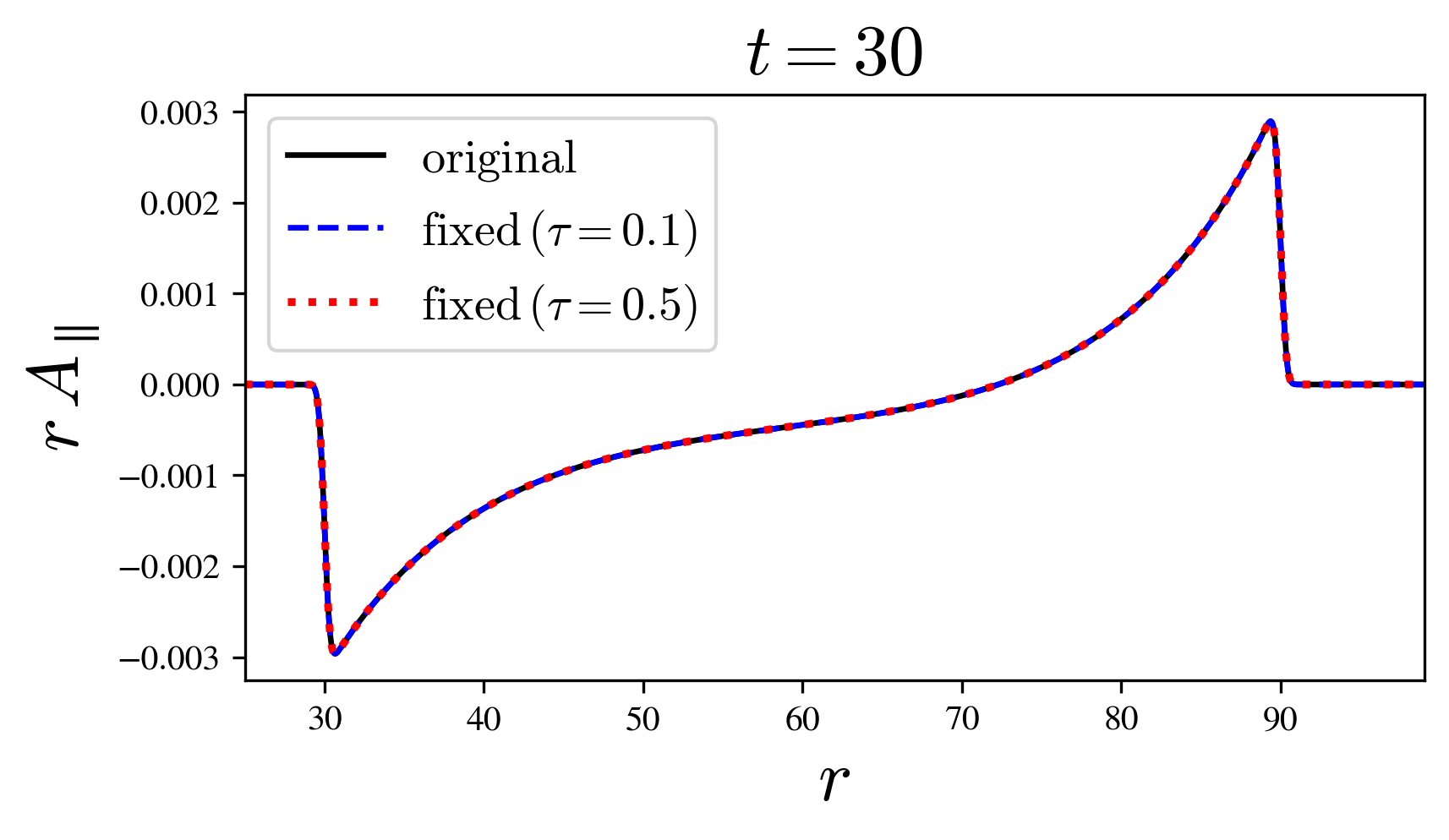}
  \includegraphics[scale=0.6]{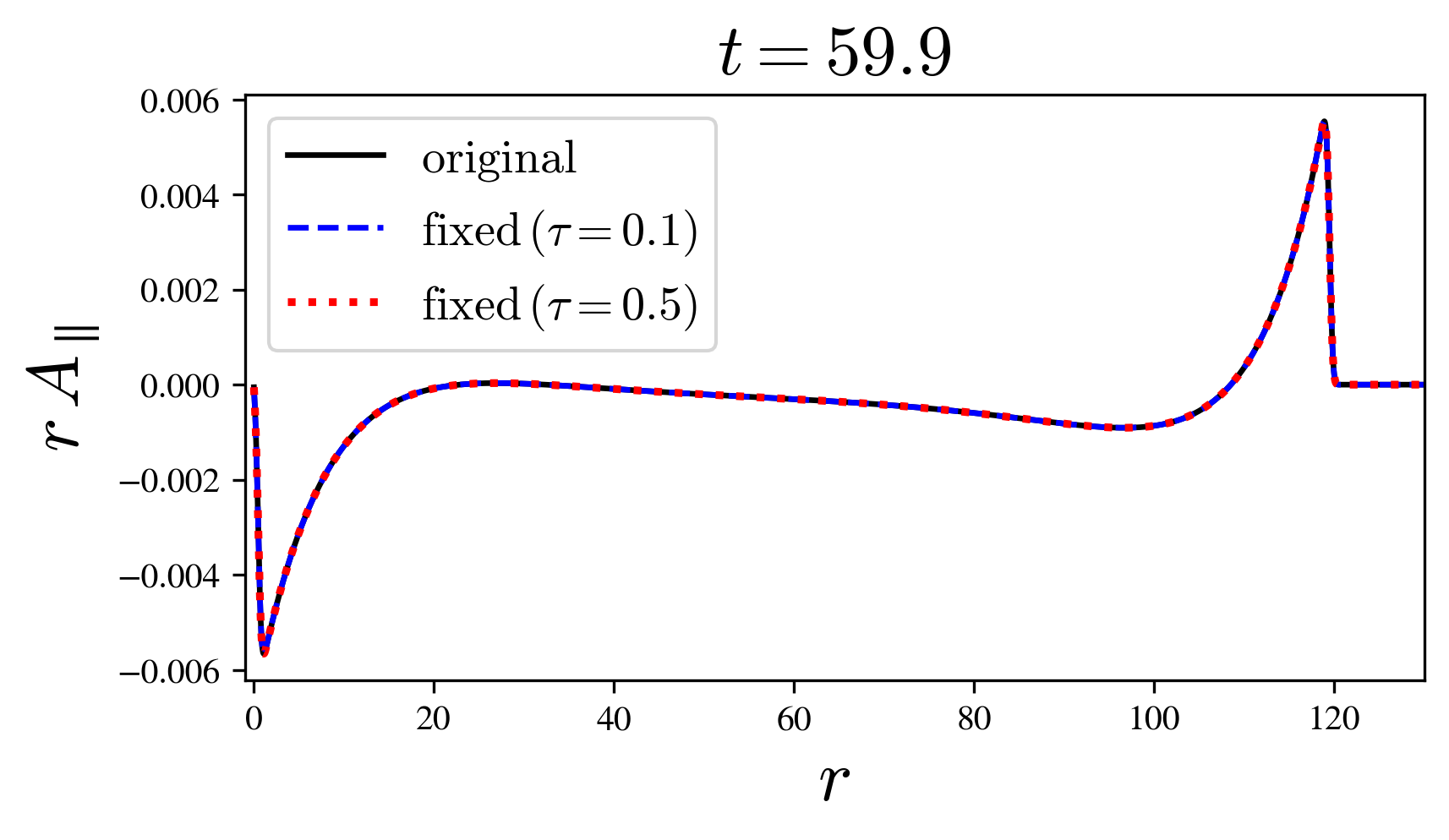}
  \includegraphics[scale=0.6]{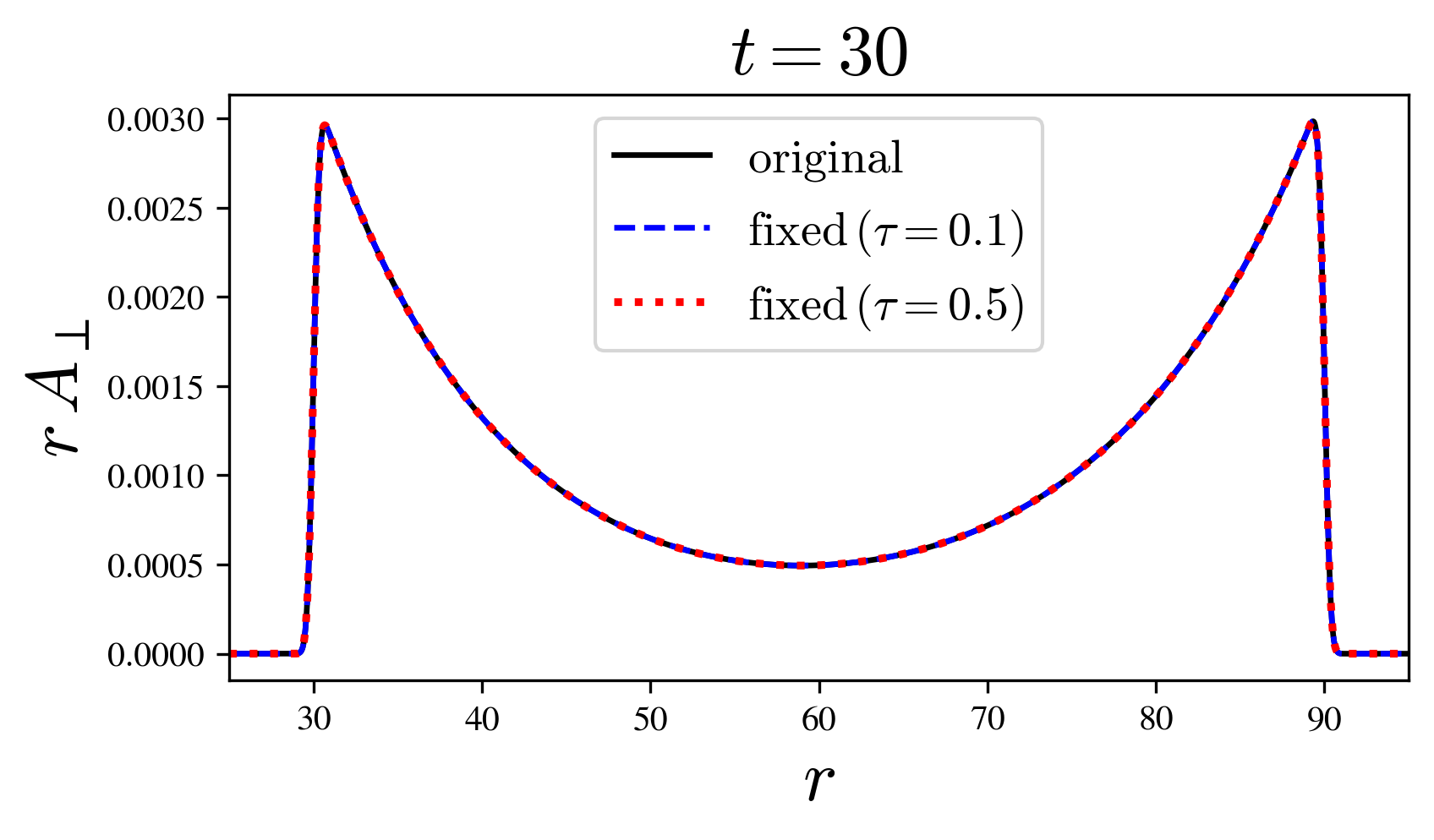}
  \includegraphics[scale=0.6]{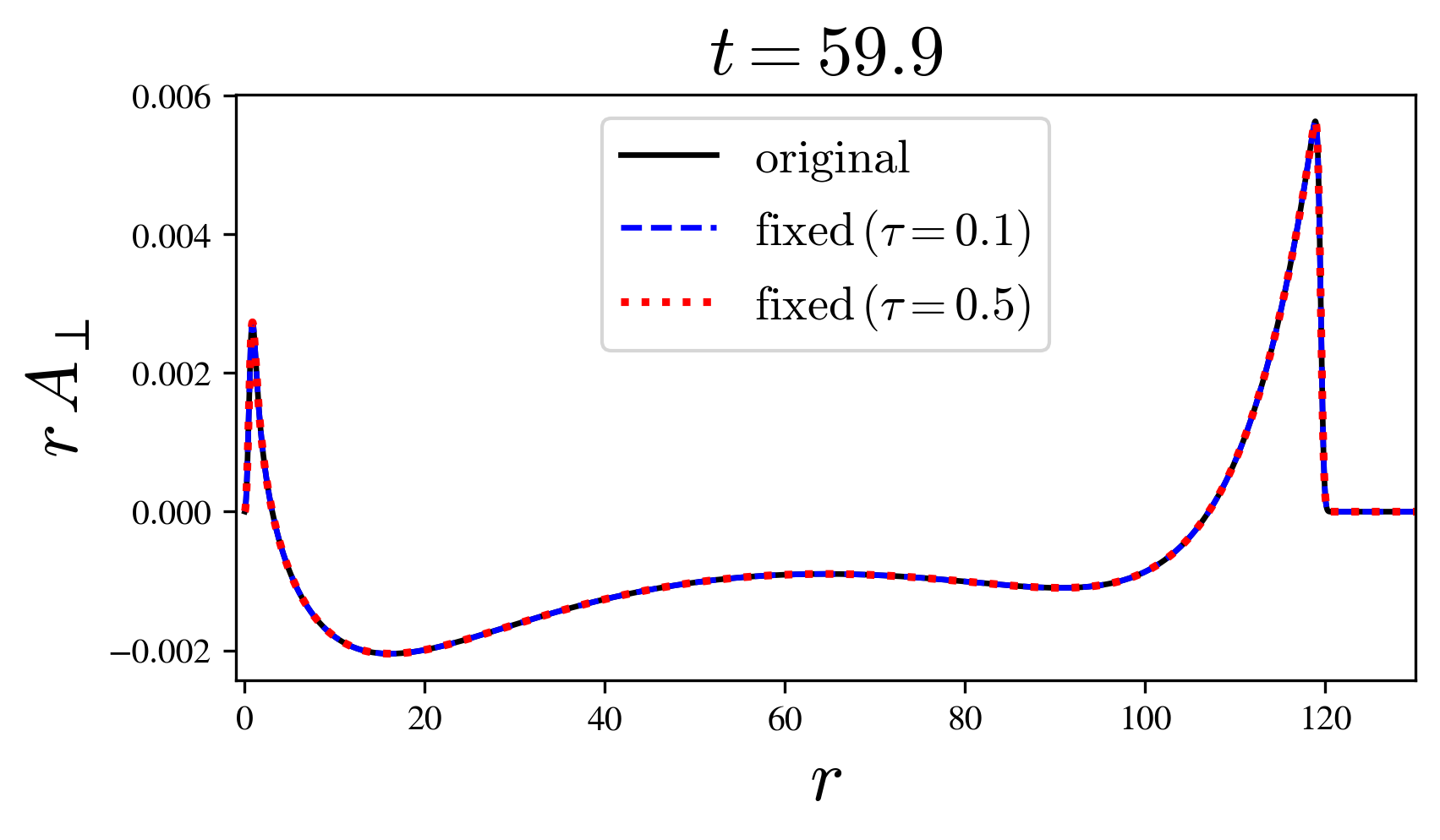}
  \caption{\textit{Dynamics of the St\"uckelberg scalar and vector fields.} Comparison between the evolutions of the theory in the original (solid black curve) and fixed versions, with parameters $\tau = 0.5$ (dotted red) and $\tau = 0.1$ (dashed blue). \textbf{Left column:} Derivative of the scalar field (top); time and radial components of the vector field (mid and bottom) at $t = 30$. \textbf{Right column:} same fields, but at $t = 59.9$ in which the Tricomi-type breakdown holds. Although a small discrepancy in the scalar profiles is observed between the approximations given by the fixed solutions, the profiles of the components of the vector remain almost unaltered. The parameters of the initial pulse for the radial derivative of the scalar field are $A = 10^{-4}$, $\sigma=0.4$ and $r_c = 60$.}
  \label{q-dynamics}
\end{figure*}

\section{Results}
\label{sec-Results}

In this section, we show the results of our simulations. For computational convenience, we employ code units such that $c=1$ and $\ell_m = t_m = m^{-2}\kappa^{-1/2}$ for length and time, where $\kappa = 8\pi G$. 

We take $\Lambda=100$ for all the simulations, $m=0.1$ in subsections \ref{subsec-tricomi} and \ref{subsec-cubic}, and $m=1$ in subsection \ref{subsec-collapse}, in the above units.

\subsection{Fixing a Tricomi-type breakdown}
\label{subsec-tricomi}

We present an example of the dynamical evolution of the Proca field in the St\"uckelberg formulation. We take $\beta=1$ and $\gamma=0$ for the coupling constants. For this choice, a Tricomi-type breakdown of the strong hyperbolicity is expected (for a wide variety of initial data) from previous studies in \textit{k}-essence \cite{Bernard19,Bezares_Kdynamics2020,MemoUV2022}. In fact, this is exactly the example case where pathologies were initially reported in \cite{Proca_CloughPRL2022, Proca_Coates1_PRL2022, Zong-Gang22}.
We adopt Type I initial data, which corresponds to a localized Gaussian pulse for the radial derivative of the scalar field, with amplitude $A = 10^{-4}$, width $\sigma=0.4$ and centered at $r_c = 60$. The time and radial components of the vector potential, as well as the radial component of the electric field, are all set initially to zero. In particular, the Gauss constraint (\ref{gauss}) is initially exactly satisfied. For the numerical simulations, we consider a radial domain of length $L=300$, a spatial resolution $\Delta r = 0.01$, and a CFL factor $C = 0.25$, so that $\Delta t = C \Delta r$.

Figure \ref{q-dynamics} shows a comparison of the dynamics yielded by the original theory (solid black) and the fixed one, with two values of the timescale; $\tau = 0.5$ (doted red) and $\tau = 0.1$ (dashed blue). The profiles of the scalar and vector fields are displayed at two different times. The first column corresponds to $t=30$, and represents initial stages of the evolution; i.e., far from the hyperbolicity breakdown. As expected, the profiles corresponding to the original and fixed theories perfectly agree, showing the robustness of the fixing, at least during the initial evolution. The second column, on the other hand, displays the dynamics of the fields at (approximately) the time in which a Tricomi-type breakdown occurs, which is $t_{\scriptsize{\mbox{tricomi}}}\sim 60$. A small discrepancy is observed in the scalar profile, between the fixed and original evolutions, while no major discrepancies are reported for the vector profiles. It is expected that the fixing may give rise to discrepancies in the evolutions close to the time in which the breakdown of hyperbolicity happens. Nevertheless, we can assess the robustness of this approximation by noticing that, as $\tau$ decreases, the fixed evolution becomes closer to the physical solution. This trend is shown in the top panel of the second column.

\begin{figure}[h]
    \centering
  \includegraphics[scale=0.72]{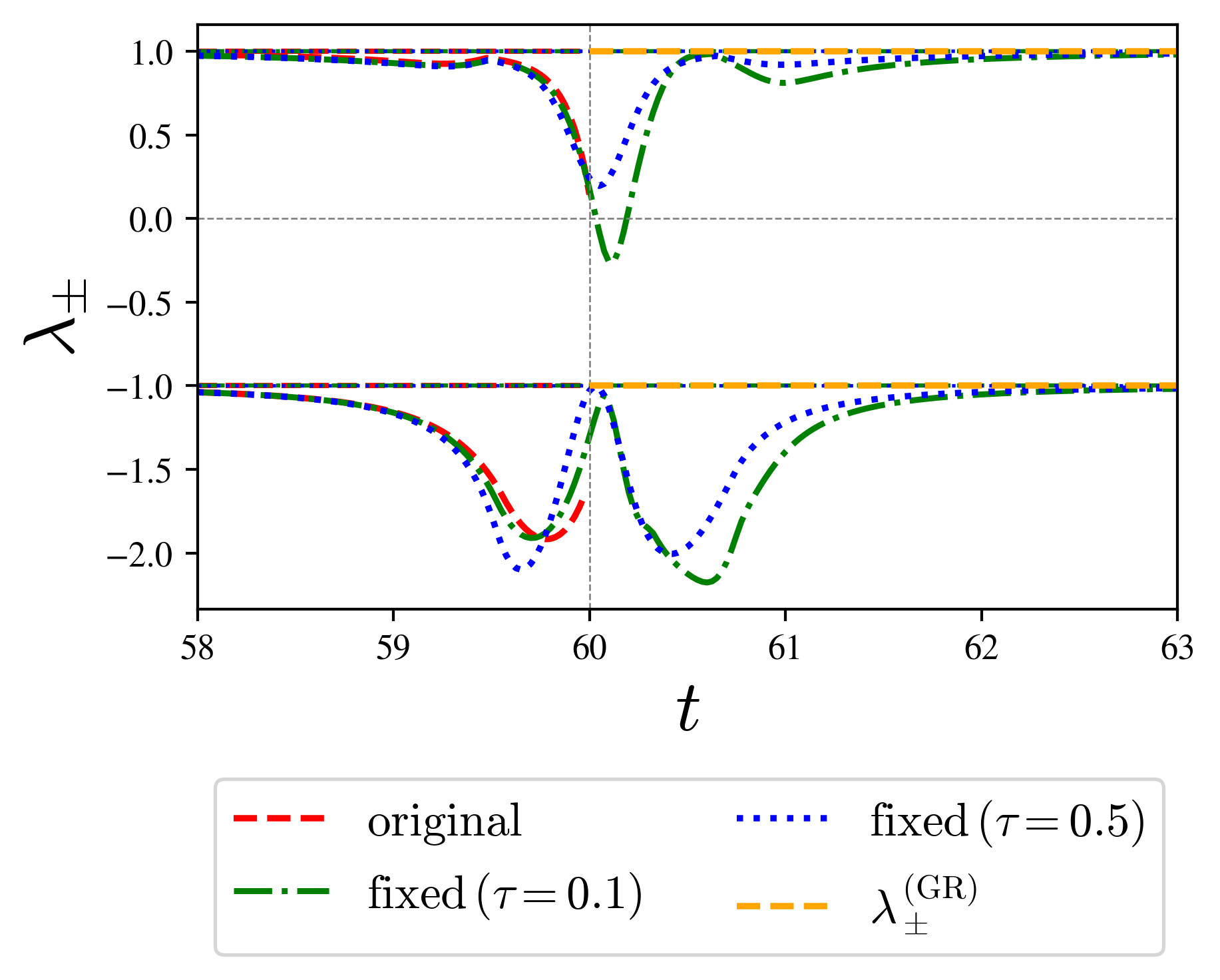}
  \caption{\textit{Eigenvalues of the effective metric.}  Minimum and maximum of the eigenvalues of $\gamma^{\mu\nu}$ for the original (dashed red) and fixed theories, with $\tau=0.1$ (dotted-dashed green) and $\tau=0.5$ (dotted blue). A Tricomi-type breakdown is observed at approximately $t_{\scriptsize{\mbox{tricomi}}} \sim 60$, for which $\lambda_+\to 0$ (upper red). The dotted blue and dashed-dotted green curves approach the eigenvalues of the fixed equations (dashed orange) after a short transient time.}
  \label{q-eigen}
\end{figure}

Figure \ref{q-eigen} exhibits the evolution of the minimum and maximum of the eigenvalues $\lambda_\pm$ of the effective metric $\gamma^{\mu\nu}$ governing the dynamics of the scalar field. In analogy with Figure \ref{q-dynamics}, we compare the evolution of the original theory with the fixed approximations, taking the same values for the parameter $\tau$ as before. The dashed red curve corresponds to the evolution of the original theory, which ceases at approximately $t\sim t_{\scriptsize{\mbox{tricomi}}}$; i.e., where the breakdown of hyperbolicity occurs. 
In particular, we observe that $\lambda_+\to 0$, meaning a Tricomi-type behavior. In order to verify that this is actually the case, the evolution is followed up with more time resolution, by decreasing the CFL factor. 
On the other hand, the dotted blue and dotted-dashed green curves correspond to the evolutions with the fixed equations. Once again, an agreement on the behavior of the eigenvalues is observed until a few time steps before the breakdown takes place, while a better approximation is reached for the smaller value of the timescale (i.e., for $\tau=0.1$). The lapse evolution was also monitored, observing that its value remains very close to unity until $t_{\scriptsize{\mbox{tricomi}}}$. This shows that, even for a weak initial pulse, a Tricomi-like breakdown of hyperbolicity can arise. 

For times greater than $t_{\scriptsize{\mbox{tricomi}}}$, the evolution continues if followed with
the fixed equations (\ref{fixed-scalar}). Now, the characteristic structure of the (fixed) scalar evolution corresponds to the one for the wave equation in GR, yielding the eigenvalues
\begin{equation}\label{fixed-eigenvalues}
    \lambda^{\tiny{\mbox{(GR)}}}_+ = \frac{1}{g_{rr}}\,, \quad \lambda^{\tiny{\mbox{(GR)}}}_- = -\frac{1}{\alpha^2}\,.
\end{equation}
After a short transitional time once the Tricomi-type breakdown has taken place in the original theory, the prediction for the minimum and maximum of the eigenvalues provided by the fixed evolution approaches their expected behaviours (dashed orange lines).

\begin{figure*}
  \centering
  \includegraphics[scale=0.75]{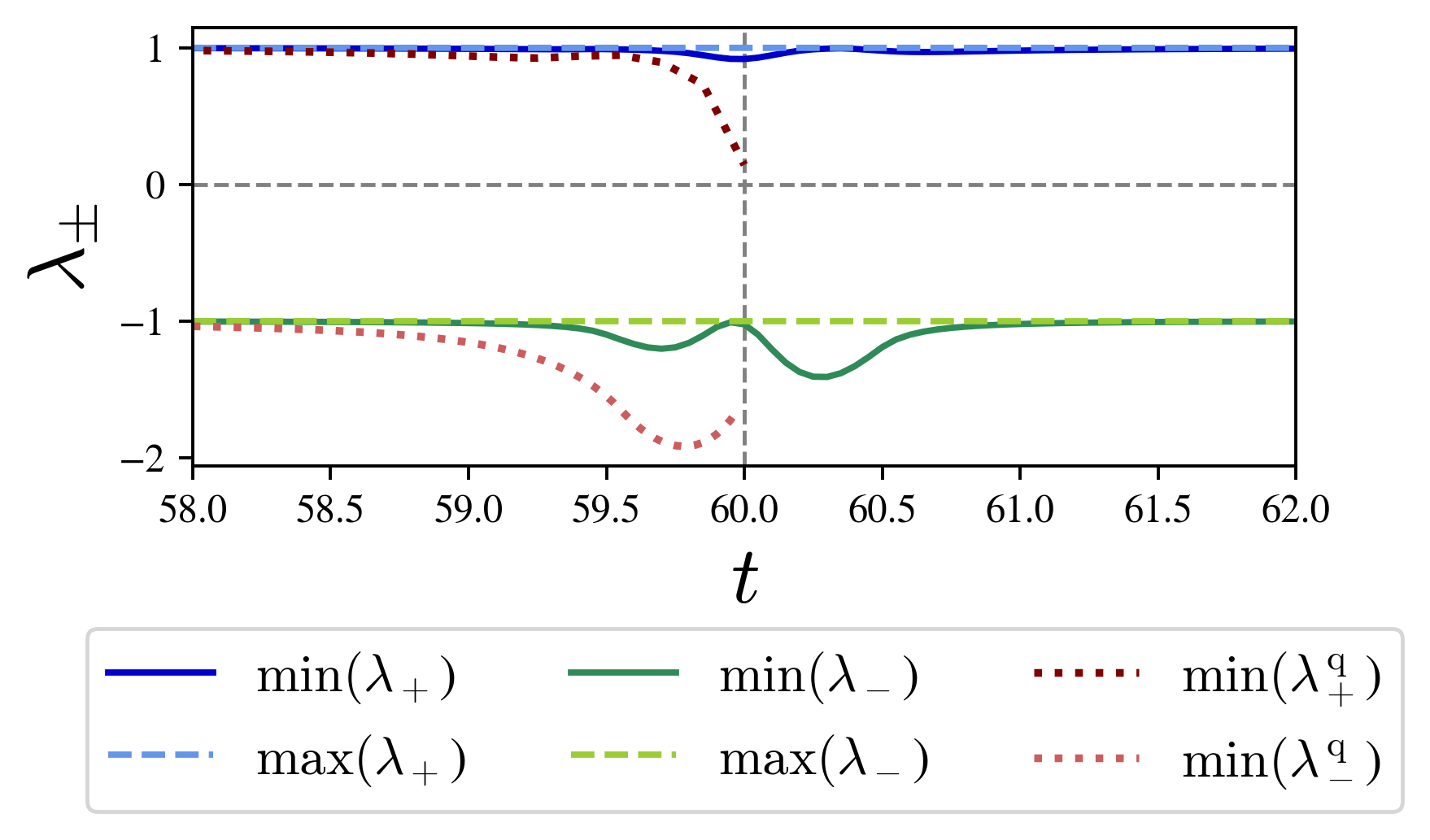}
  \includegraphics[scale=0.75]{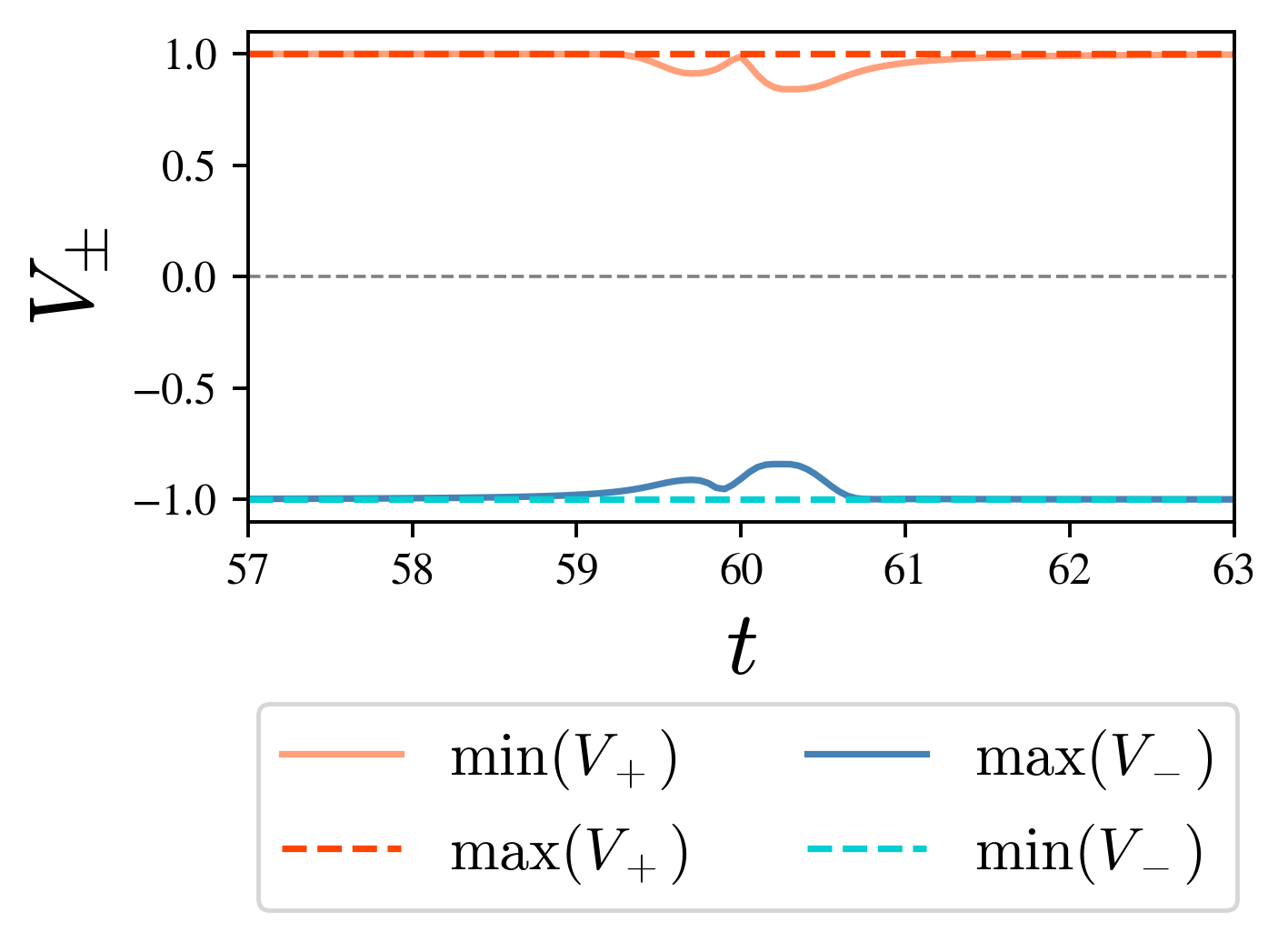}
  \caption{\textit{Characteristic evolution with a cubic self-interaction.}  \textbf{Left panel:} maximum and minimum of the eigenvalues of the effective metric as a function of time. From top to bottom: $\max(\lambda_+)$ (dashed blue); $\min(\lambda_+)$ (solid blue); $\max(\lambda_-)$ (dashed green) and $\min(\lambda_-)$ (solid green). A comparison with the case with only a quadratic self-interaction is shown in dashed brown for the minimum of the corresponding eigenvalues $\lambda_\pm^{\text{q}}$. \textbf{Right panel:} maximum and minimum of the characteristic velocities $V_\pm$ of the scalar evolution system. From top to bottom: $\max(V_+)$ (dashed orange); $\min(V_+)$ (solid orange); $\max(V_-)$ (solid blue) and $\min(V_-)$ (dashed turquoise).}
  \label{fig:cubic-charac}
\end{figure*}

\subsection{Evolution with a cubic self-interaction}
\label{subsec-cubic}

We also study the evolution of the Proca field with a cubic self-interaction, taking $\beta=\gamma=1$ for the couplings. We evolve an initial data set very similar to the one considered in the previous section. Although we set the same parameters for the Gaussian pulse, the elliptic equation (\ref{HamConstraint}) needed to obtain the initial metric fields depends on $K(X)$, which now includes a cubic term. The numerical domain, spatial resolution and CFL factor are taken as in the simulation showed before. 
From Eq. (\ref{det-gamma}), it is straightforward to see that, for this choice of the coupling constants, there can not be a breakdown of hyperbolicity of Tricomi-type, since $\det(\gamma^{\mu\nu}) < 0$ for all $X$, in agreement with \cite{ShortPaper_Barausse2022,Bezares_Kdynamics2020}.
We have confirmed this fact numerically, i.e. a Tricomi-type failure, which can arise from the action considered in \cite{Proca_CloughPRL2022, Proca_Coates1_PRL2022, Zong-Gang22}, can be easily avoided by adding a cubic term.
Moreover, we numerically find that a Keldysh-type behavior does not occur either.

\begin{figure}[h]
    \centering
  \includegraphics[scale=0.61]{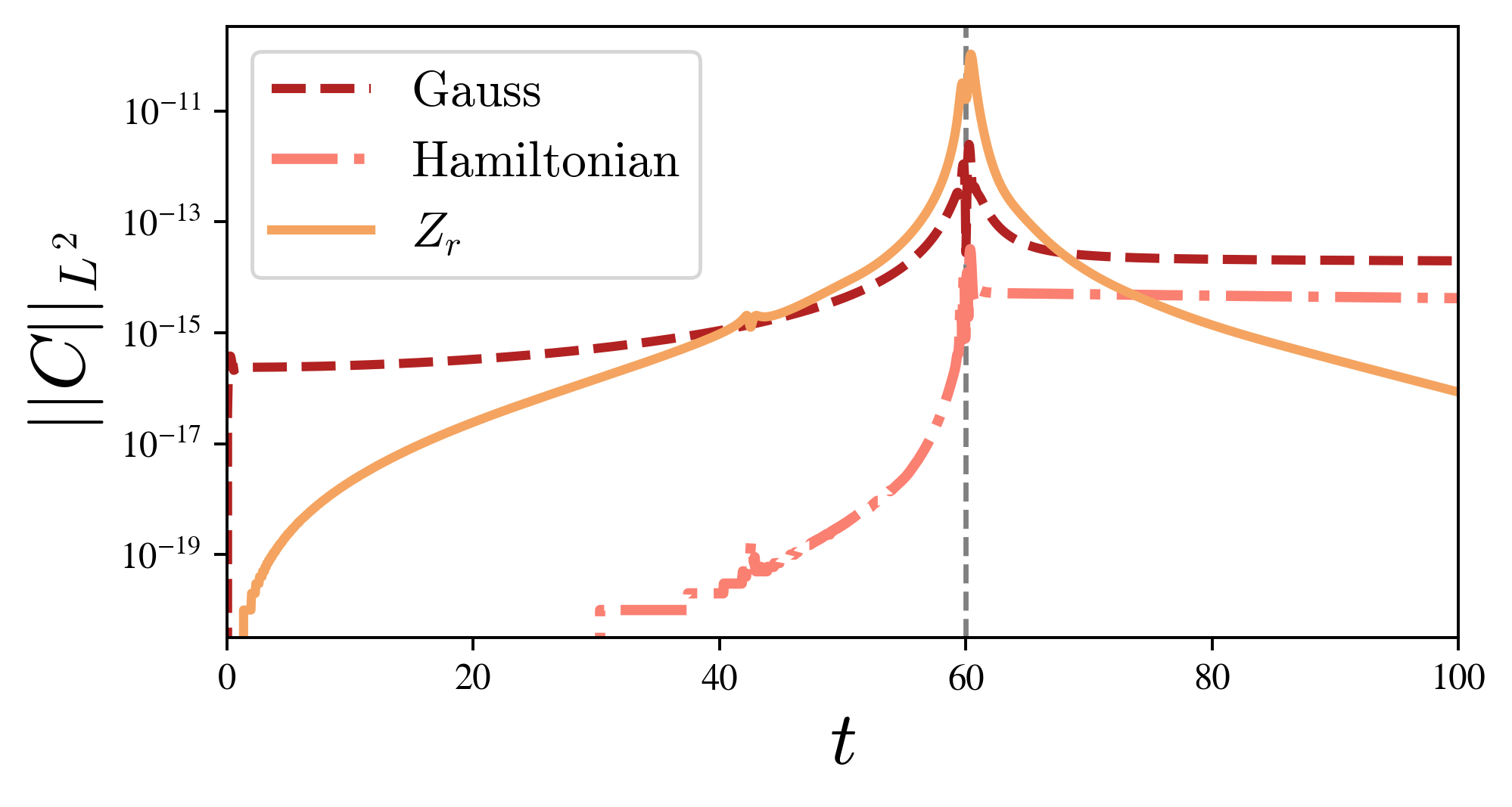}
  \caption{\textit{Constraint propagation.} $L^2$ norm of the Gauss (dashed brown), Hamiltonian (dashed-dotted salmon) and Z3 (solid orange) constraints as a function of time, when considering a cubic self-interaction.}
  \label{fig:cubic-constraints}
\end{figure}

Figure \ref{fig:cubic-charac} illustrates essential aspects of the characteristic structure of the evolution system. The left panel represents the maximum and minimum values of the eigenvalues of the effective metric $\gamma^{\mu\nu}$ as a function of time (light-blue dashed and blue continuous lines for, respectively, the maximum and minimum of $\lambda_+$; light-green dashed and green continuous lines for, respectively, the maximum and minimum of $\lambda_-$). Also, and for illustrative purposes only, a comparison with the case $\gamma=0$ is included (brown dashed curves). The eigenvalues remain far from zero, exhibiting a well-posed and stable evolution. The right panel displays the maximum and minimum values of the characteristic velocities of the scalar equation in (\ref{full-eqs}), verifying that no Keldysh divergence of the characteristic speeds was found during the whole evolution.

The time evolution of the norm of the Gauss (dashed brown), Hamiltonian (dashed orange) and Z3 (dotted-dashed salmon) constraints is displayed in Figure \ref{fig:cubic-constraints}. In particular, they remain less than $10^{-14}$, up to $t\sim 60$, where the pulse reaches $r=0$. For that value of time, the norm grows to approximately $10^{-11}$, which is expected from the boundary/regularity conditions considered. After that, although the Z3 constraint considerably decays, the Hamiltonian and Gaussian constraints remain bounded around $10^{-14}$.

\subsection{Gravitational collapse with a cubic self-interaction}
\label{subsec-collapse}

We report here an example of the dynamical evolution of the Proca field with a cubic self-interaction leading to gravitational collapse, particularly in the formation of a black hole. For the coupling constants we take $\beta=0$ and $\gamma = 1$, afresh preventing the determinant (\ref{det-gamma}) from ever becoming zero.

We considered the Type II initial data given by Eq. (\ref{ID-typeII}), with amplitude $A = 0.093$, width $\sigma = 0.942$ and center $r_c = 55$. This configuration yielded an Arnowitt-Deser-Misner (ADM) mass $M_{\tiny{\mbox{ADM}}} \simeq 1.590$. At the first stages of the simulation, the scalar pulse splits into two modes: one moving towards the origin with an amplitude that increases, approximately, as $1/r$ (as expected) and the other towards the outer boundary. The spatial resolution, the CFL factor and the size of the numerical domain were all taken to be the same as in the cases presented earlier. 
\begin{figure}[h]
    \centering
  \includegraphics[scale=0.68]{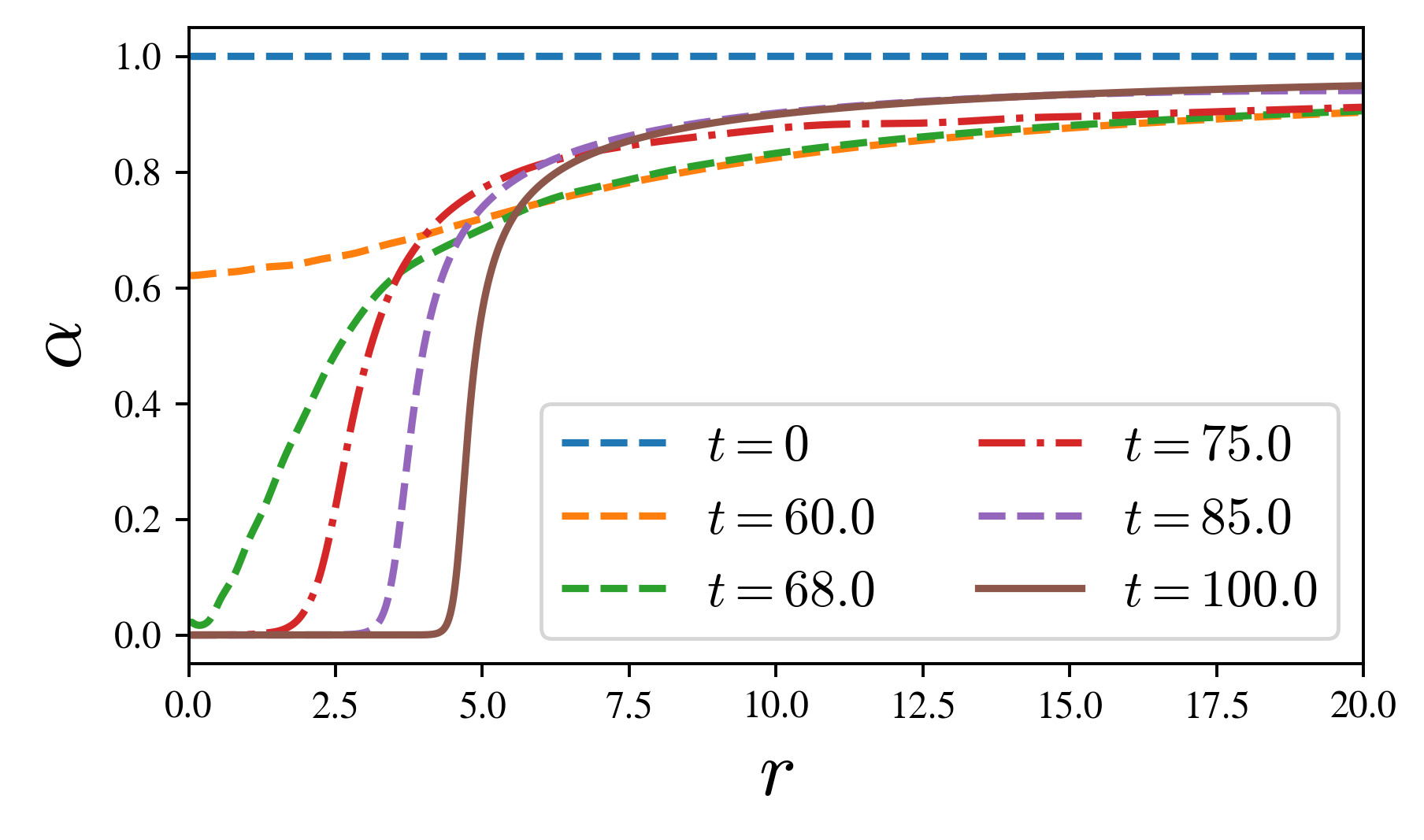}
  \caption{\textit{Gravitational collapse with a cubic self-interaction.} Profiles of the lapse function for different values of time. Although it is initially set to one, the ``$1+\log$'' slicing condition forces the lapse to vanish close to the origin. The collapse occurs around $t_{\tiny{\mbox{AH}}}\sim 68.5$, where the first apparent horizon is formed, and the lapse reaches zero soon after $t_{\tiny{\mbox{AH}}}$.}
  \label{fig:alpha-snapshots-collapse}
\end{figure}

\begin{figure}[h]
    \centering
  \includegraphics[scale=0.7]{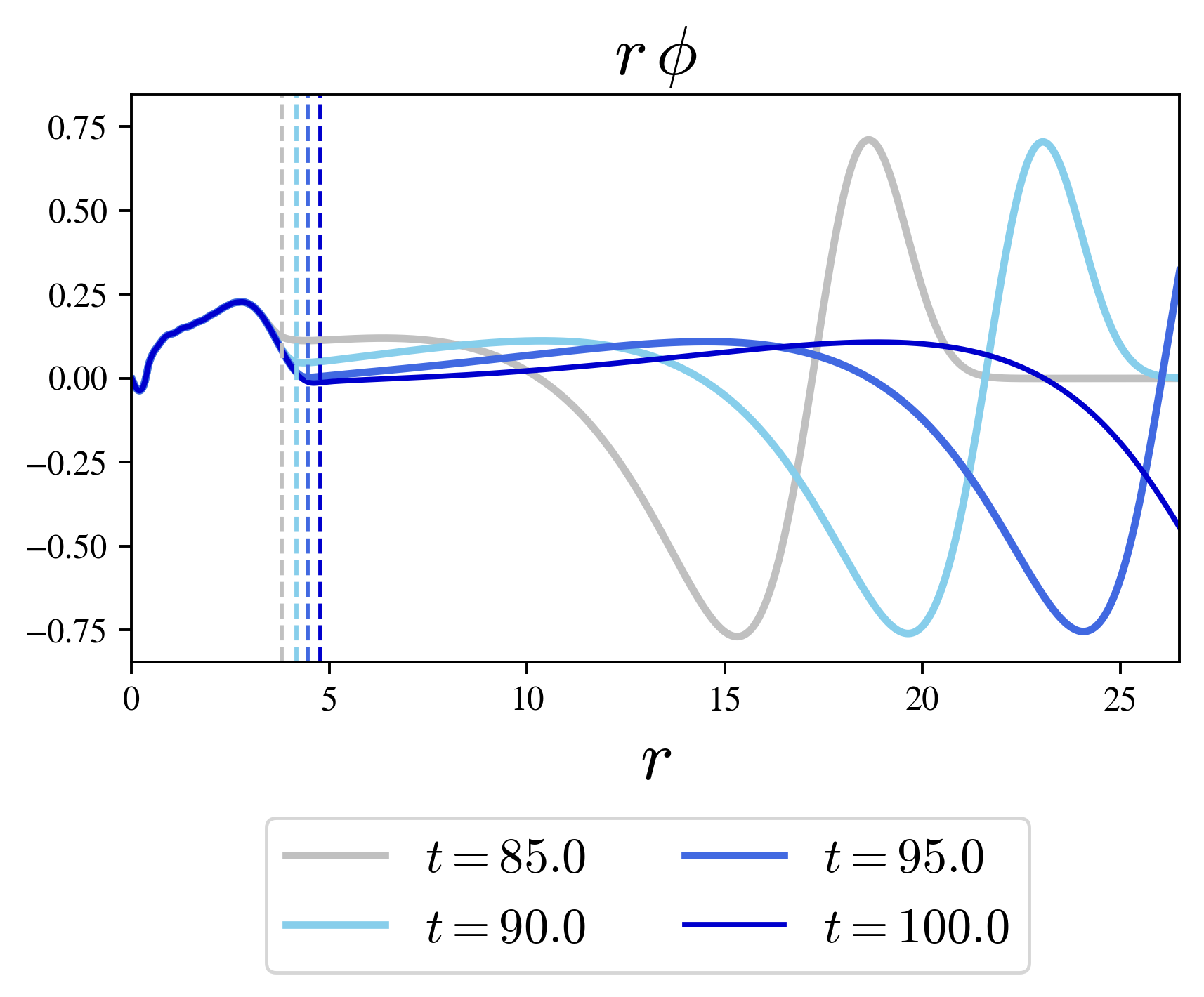}
  \includegraphics[scale=0.67]{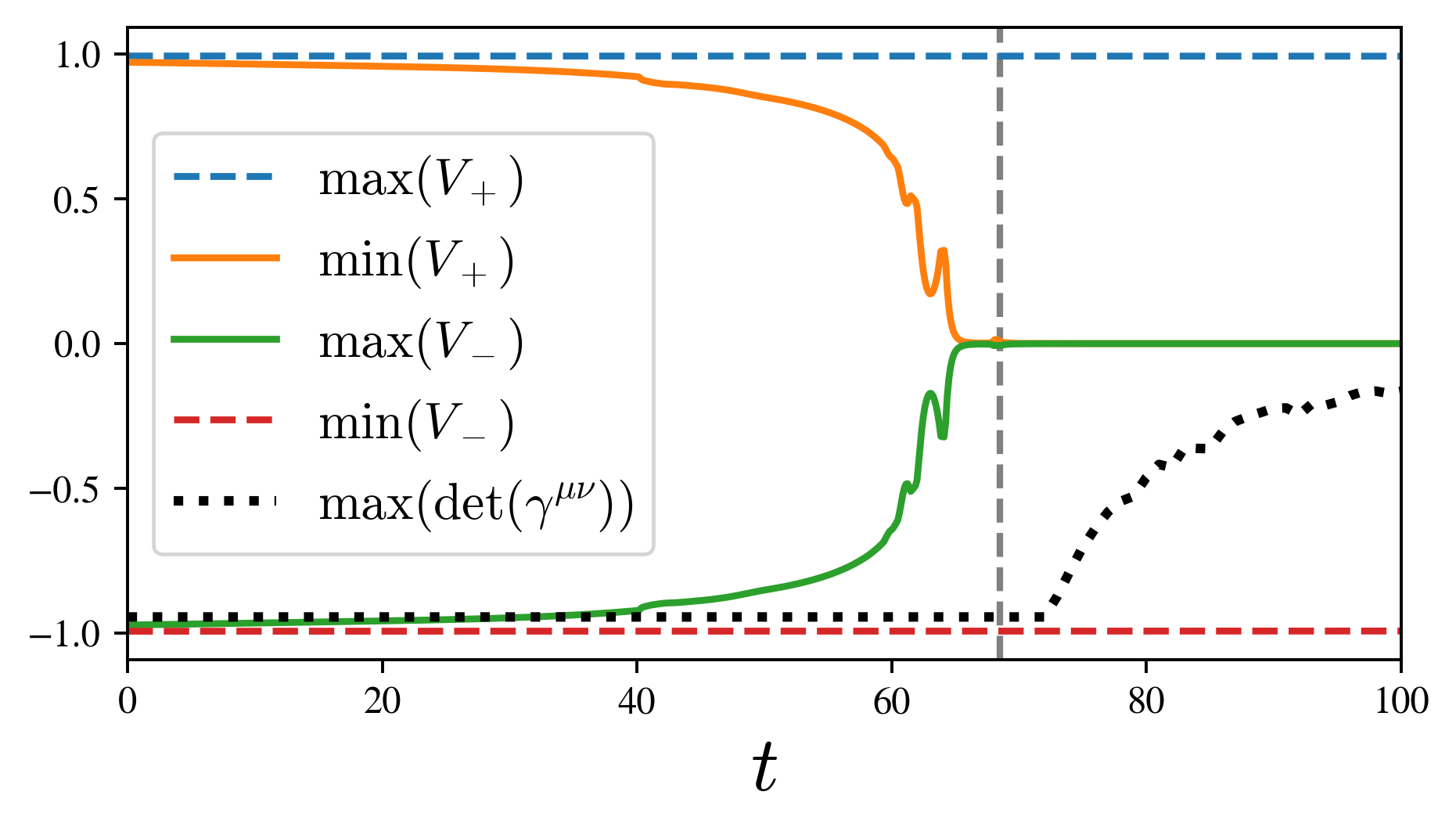}
  \caption{\textit{Dynamics of the collapse.} \textbf{Top:} Snapshots of the scalar field (multiplied by $r$) in the region close to the apparent horizon, for different times after the collapse takes place. The location of the apparent horizon at each time is shown with a vertical line. \textbf{Bottom:} minimum and maximum values of the characteristic velocities $V_\pm$ of the scalar equation as a function of time. From top to bottom: $\max(V_+)$ (dashed blue); $\min(V_+)$ (solid orange); $\max(V_-)$ (solid green) and $\min(V_-)$ (dashed red). The maximum of the determinant of the effective metric $\gamma^{\mu\nu}$ is represented in dotted black, and the time at which the first apparent horizon is formed is meant by the dashed grey vertical line.}
  \label{fig:collapse-cubic}
\end{figure}

By letting the system evolve until $t_{\tiny{\mbox{final}}}\sim 100$, the scalar field collapses and a black hole forms. This is signalled by the formation of an apparent horizon (i.e., the outermost trapped surface) at $t\sim 68.5$, whose position continuously increases in our coordinates. To keep track of the dynamics of this surface, we considered the largest value of $r$ for which the expansion of the outgoing null-ray congruence vanishes, physically indicating a confinement of the latter. In spherical symmetry, this condition reads~\cite{Thornburg:2006zb}
\begin{equation}
    D_{r\theta}{}^\theta+\frac{1}{r} - \sqrt{g_{rr}} K_\theta{}^\theta = 0,
\end{equation}
where the function $D_{r\theta}{}^\theta$ is defined in Eq. (\ref{first-order-vars}). 
The results for the lapse evolution are presented in Figure \ref{fig:alpha-snapshots-collapse}, showing profiles at different times. Starting from $\alpha=1$, the lapse vanishes at approximately $t\sim 69$, and the front propagation speed grows in time due to our gauge choice. Also, in this particular gauge the singularity is avoided, allowing one to proceed with the simulations after the appearance of the first apparent horizon. The areal radius $r_{\tiny{\mbox{A}}} = r\,\sqrt{g_{\theta\theta}}$ is also computed during the evolution. We estimate a black-hole area of $A_{\tiny{\mbox{BH}}}\simeq 13.854$, and a mass of $M_{\tiny{\mbox{BH}}} \simeq 0.525$. The latter represents approximately $33\%$ of the total ADM mass of the system. This apparent mass loss is consistent with the initial configuration: just one half of the total scalar energy of the initial pulse takes part of the physical process leading to collapse, while the other half gets rapidly radiated away.

The top panel of Figure \ref{fig:collapse-cubic} shows profiles of the St\"uckelberg field close to the horizon for different time values after the black hole has formed. The dashed vertical lines show the location of the apparent horizon at these times, which are meant by different colors. In the bottom panel, we display the maximum and minimum values of the characteristic speeds of the scalar equation, as well as the maximum of the determinant of the effective metric (dotted black). The latter remains always bounded and different from zero, assuring no breakdown of the Cauchy problem during evolution. After the black hole has formed, our gauge choice ``freezes-out'' the region within the apparent horizon, thus not giving any physical insights on the dynamics at the interior. Alternative and more sophisticated numerical  approaches 
to horizon formation would also be possible; e.g., excision of the interior region from the numerical domain (at the cost of putting boundary conditions at the horizon, which moves during the evolution), or another gauge choice. 


\section{Conclusions}
\label{sec-Conclusions}

In this paper, we conducted numerical simulations of self-interacting massive vector fields coupled to General Relativity, in spherical symmetry. We investigated the hyperbolicity of the corresponding Cauchy problem, displaying numerical evidences that the breakdowns faced by the theory are analogous to those occurring in scalar-tensor theories with first-order derivative self-interactions (usually known as \textit{k}-essence theories). 

In particular, we gave an explicit example leading to a Tricomi-type breakdown, when only quadratic self-interactions are accounted for. We showed that this pathology can be avoided by suitably deforming the evolution equations by means of a ``fixing-the-equations'' approach. 
We also explored the dynamics of the vector field when cubic self-interactions are taken into account. As expected, no Tricomi-type breakdown of the Cauchy problem occurred.
For the latter case, we also explored gravitational collapse. We found suitable initial data giving rise to well-posed and stable evolutions towards a black hole final state.

To conclude, we stress the advantage of our approach for performing numerical simulations of massive vector fields, as it singles out the St\"uckelberg mode, taming derivative self-interactions. In particular, it can also be implemented in massive gravity (i.e., a massive spin-2 field breaking diffeomorphism invariance, which can be restored by a St\"uckelberg transformation), with potential applications to the study of instabilities around black holes, as done in \cite{East:2023nsk}, or in the context of black hole superradiance \cite{Brito:2013wya,brito2020superradiance}. We leave these explorations for future work.


\section*{Acknowledgements}

We thank Ramiro Cayuso, Luis Lehner and Marc Schneider for discussions throughout this work.
MR acknowledges hospitality from the Perimeter Institute for Theoretical Physics, where part of this work was carried out.
MR and EB acknowledge support from the European Union’s H2020 ERC Consolidator Grant ``GRavity from Astrophysical to Microscopic Scales'' (Grant No. GRAMS-815673), the PRIN 2022 grant ``GUVIRP - Gravity tests in the UltraViolet and InfraRed with Pulsar timing'', and the EU Horizon 2020 Research and Innovation Programme under the Marie Sklodowska-Curie Grant Agreement No. 101007855.
MC is funded by the European Union under the Horizon Europe's Marie Sklodowska-Curie project~101065440.

\appendix

\bibliographystyle{ieeetr}
\bibliography{proca}
	
\end{document}